\documentclass[aps,prl,preprint,nopacs,superscriptaddress, font]{revtex4}
\Large
\usepackage{graphicx}
\usepackage{verbatim}
\usepackage{mathrsfs}
\usepackage{amsmath,amsfonts,amssymb}
\usepackage{graphicx}
\def\3{2.8in}    %used for figure widths
\def\2{2.5in}
\def\4{3.0in}

\def \beq {\begin{equation}}
\def \eeq {\end{equation}}
\begin{document}

\title{Discovery of a three-dimensional topological Dirac semimetal phase in high-mobility Cd$_3$As$_2$}

%Observation of a three-dimensional topological Dirac semimetal phase in high-mobility Cd$_3$As$_2$

\author{Madhab Neupane*}\affiliation {Joseph Henry Laboratory, Department of Physics, Princeton University, Princeton, New Jersey 08544, USA}

\author{Su-Yang Xu*}\affiliation {Joseph Henry Laboratory, Department of Physics, Princeton University, Princeton, New Jersey 08544, USA}

\author{Raman Sankar*} \affiliation{Center for Condensed Matter Sciences, National Taiwan University, Taipei 10617, Taiwan}

\author{Nasser Alidoust}\affiliation {Joseph Henry Laboratory, Department of Physics, Princeton University, Princeton, New Jersey 08544, USA}

\author{Guang Bian}\affiliation {Joseph Henry Laboratory, Department of Physics, Princeton University, Princeton, New Jersey 08544, USA}

\author{Chang Liu}\affiliation {Joseph Henry Laboratory, Department of Physics, Princeton University, Princeton, New Jersey 08544, USA}

\author{Ilya Belopolski}\affiliation {Joseph Henry Laboratory, Department of Physics, Princeton University, Princeton, New Jersey 08544, USA}

\author{Tay-Rong Chang} \affiliation{Department of Physics, National Tsing Hua University, Hsinchu 30013, Taiwan}

\author{Horng-Tay Jeng} \affiliation{Department of Physics, National Tsing Hua University, Hsinchu 30013, Taiwan} \affiliation{Institute of Physics, Academia Sinica, Taipei 11529, Taiwan}

\author{Hsin Lin}\affiliation {Graphene Research Centre and Department of Physics,
National University of Singapore, Singapore 117542}

\author{Arun Bansil}\affiliation {Department of Physics, Northeastern University, Boston, Massachusetts 02115, USA}

\author{Fangcheng Chou} \affiliation{Center for Condensed Matter Sciences, National Taiwan University, Taipei 10617, Taiwan}

\author{M. Zahid Hasan}\affiliation {Joseph Henry Laboratory, Department of Physics, Princeton University, Princeton, New Jersey 08544, USA}
\affiliation {Princeton Center for Complex Materials, Princeton University, Princeton, New Jersey 08544, USA}


%\newpage
\begin{abstract}

\textbf{Symmetry-broken three-dimensional topological Dirac semimetal systems with strong spin-orbit coupling can host many exotic Hall-like phenomena and Weyl Fermion quantum transport.
Here using high-resolution angle-resolved photoemission spectroscopy, we performed systematic electronic structure studies on Cd$_3$As$_2$, which has been predicted to be the parent material, from which many unusual topological phases can be derived. We observe a highly linear bulk band crossing to form a three-dimensional dispersive Dirac cone projected at the Brillouin zone center by studying the (001)-cleaved surface. Remarkably, an unusually in-plane high Fermi velocity up to 1.5 $\times$ 10$^{6}$ ms$^{-1}$ is observed in our samples, where the mobility is known up to 40,000 cm$^2$V$^{-1}$s$^{-1}$ suggesting that Cd$_3$As$_2$ can be a promising candidate as an anisotropic-hypercone (3D) high spin-orbit analog of graphene. Our experimental identification of the Dirac-like bulk topological semimetal phase in Cd$_2$As$_2$ opens the door for exploring higher dimensional spin- orbit Dirac physics in a real material.}

%m$\cdot$s$^{-1}$

%\textbf{Experimental realization of two dimensional Dirac systems has recently led to a flurry of research activity in condensed matter physics. Symmetry-broken three dimensional (3D) Dirac systems with strong spin-orbit coupling can also host many exotic Hall-like phenomena and quantum transport such as the Weyl phases, linear quantum magnetoresistance and topological magnetic phases. Using high resolution angle-resolved photoemission spectroscopy, we performed systematic electronic structure studies on Cd$_3$As$_2$, which has been predicted to be the parent material, from which many unusual topological phases can be derived. We observe a highly linear bulk band crossing to form a three-dimensional dispersive Dirac cone projected at the Brillouin zone center by studying the (001)-cleaved surface. Remarkably, an unusually in-plane high Fermi velocity (more than an order of magnitude higher than the theoretically predicted value) up to 9.8 $\textrm{\AA}{\cdot}$eV (1.5 $\times$ 10$^{6}$ m$\cdot$s$^{-1}$) is observed in our samples, where the mobility is known up to 40,000 cm$^2$V$^{-1}$s$^{-1}$ suggesting that Cd$_3$As$_2$ can be a promising candidate as an anisotropic-hypercone (3D) high spin-orbit analog of graphene. Our experimental identification and band-structure measurements of the Dirac-like bulk semimetal phase and its clear contrast with Bi$_2$Se$_3$ and 2D graphene discovered previously, opens the door for exploring higher dimensional spin-orbit Dirac physics in a stoichiometric material.}

\end{abstract}
\date{\today}
\maketitle

%\newpage
% Our photon energy dependent studies reveal band dispersions also along the \textit{out-of-plane} $k_z$ momentum space direction providing evidence for the 3D nature of the observed bands in Cd$_3$As$_2$ which is in contrast to the well studied topological insulators with anisotropic surface states as in Bi$_2$Te$_3$. Our data also show similarities between Cd$_3$As$_2$ and Bi-based 3D band-inversion-critical materials.

Two-dimensional (2D) Dirac electron systems exhibiting many exotic quantum phenomena constitute one of the most active topics in condensed matter physics \cite{Graphene, Weyl, RMP, Zhang_RMP, David_nature, Xia, Chen_Science, Hasan2, HsiehSci, Dirac_3D, Bismuth, Dirac_semi, 3D_Dirac, Dai, Neupane, Volovik, Fang, Ashvin, Balent}. The notable examples are graphene and the surface states of topological insulators (TI). Three-dimensional (3D) Dirac fermion metals, sometimes noted as the topological bulk Dirac semimetal (BDS) phases, are also of great interest if the material possesses 3D isotropic or anisotropic relativistic dispersion in the presence of strong spin-orbit coupling. It has been theoretically predicted that a topological (spin-orbit) 3D spin-orbit Dirac semimetal can be viewed as a composite of two sets of Weyl fermions where broken time-reversal or space inversion symmetry can lead to a surface Fermi-arc semimetal phase or a topological insulator \cite{Dai}. In the absence of spin-orbit coupling, topological phases cannot be derived from a 3D Dirac semimetal. Thus the parent BDS phase with strong spin-orbit coupling is of great interest. Despite their predicted existence \cite{3D_Dirac, Dirac_semi, Dai}, experimental studies on the massless BDS phase have been lacking since it has been difficult to realize this phase in real materials, especially in stoichiometric single crystalline non-metastable systems with high mobility. It has also been noted that the BDS state can be achieved at the critical point of a topological phase transition \cite{Suyang, Ando} between a normal insulator and a topological insulator which requires fine-tuning of the chemical doping/alloying composition thus by effectively varying the spin-orbit coupling strength. This approach also introduces chemical disorder into the system. In stoichiometric bulk materials, the known 3D Dirac fermions in bismuth are in fact of massive variety since there clearly exists a band gap in the bulk Dirac spectrum \cite{ Bismuth}. On the other hand, the bulk Dirac fermions in the Bi$_{1-x}$Sb$_x$ system coexist with additional Fermi surfaces \cite{David_nature}. Therefore, to this date, identification of a gapless BDS phase in stoichiometric materials remains experimentally elusive.

%can lead to peculiar physical properties \cite{Dirac_3D, Bismuth, Dirac_semi, 3D_Dirac, Dai, Volovik, Fang, Ashvin, Balent}.

In this article, we present the experimental identification of a gapless Dirac-like 3D topological (spin-orbit) semimetal phase in stoichiometric single crystalline system of Cd$_3$As$_2$, which is protected by the $C_4$ crystalline (crystal structure) symmetry and spin-orbit coupling as predicted in theory \cite{Dai}. Using high-resolution angle-resolved photoemission spectroscopy (ARPES), we show that Cd$_3$As$_2$ features a bulk band Dirac-like cone locating at the center of the (001) surface projected Brillouin zone (BZ). Remarkably, we observe that the band velocity of the bulk Dirac spectrum is as high as $\sim$ 10 $\textrm{\AA}{\cdot}$eV, which along with its massless character favorably contributes to its natural high mobility ($\sim$ $10^5$ cm$^2$V$^{-1}$s$^{-1}$ \cite{Mobi, Mobi2}). We further compare and contrast the observed crystalline-symmetry-protected BDS phase in Cd$_3$As$_2$ with those of in the Bi-based 3D-TI systems such as in BiTl(S$_{1-\delta}$Se$_{\delta}$)$_2$ and (Bi$_{1-\delta}$In$_{\delta}$)$_2$Se$_3$ systems. Our experimental identification and band-structure measurements of the Dirac-like bulk semimetal phase and its clear contrast with Bi$_2$Se$_3$ and 2D graphene discovered previously, opens the door for exploring higher dimensional spin-orbit Dirac physics in a stoichiometric material. These new directions are uniquely enabled by our observation of strongly spin-orbit coupled 3D massless Dirac semimetal phase protected by the $C_4$ symmetry, which is not possible in the 2D Dirac fermions in graphene and the surfaces of topological insulators, or weak spin-orbit 3D Dirac fermions in other materials.

\bigskip
\bigskip
\textbf{Results}
\newline

\textbf{Crystalline symmetry protected topological Dirac phase} 

The crystal structure of Cd$_3$As$_2$ has a tetragonal unit cell with $a= 12.67$ $\AA$ and $c= 25.48$ $\AA$ for $Z= 32$ with symmetry of space group $I4_1$cd (see Figs. 1a and b). In this structure, arsenic ions are approximately cubic close-packed and Cd ions are tetrahedrally coordinated, which can be described in parallel to a fluorite structure of systematic Cd/As vacancies. There are four layers per unit and the missing Cd-As$_4$ tetrahedra are arranged without the central symmetry as shown with the (001) projection view in Fig.1b, with the two vacant sites being at diagonally opposite corners of a cube face \cite{crys_str}. The corresponding Brillouin zone (BZ) is shown in Fig. 1d, where the center of the BZ is the $\Gamma$ point, the centers of the top and bottom square surfaces are the $Z$ points, and other high symmetry points are also noted. Cd$_3$As$_2$ has attracted attention in electrical transport due to its high mobility of $10^5$ cm$^2$V$^{-1}$s$^{-1}$ reported in previous studies \cite{Mobi, Mobi2}. The carrier density and mobility of our Cd$_3$As$_2$ samples (shown in Fig. 1 and 2) are characterized to be of $5.2\times10^{18}$ cm$^{-3}$ and $42850$ cm$^2$V$^{-1}$s$^{-1}$, respectively, at temperature of 130 K, consistent with previous reports \cite{Mobi, Mobi2}, which provide an evidence for the high quality of our single crystalline samples. In band theoretical calculations, Cd$_3$As$_2$ is also of interest since it features an inverted band structure \cite{Inverted}. More interestingly, a very recent theoretical prediction \cite{Dai} which motivated this work, has shown that the spin-orbit interaction in Cd$_3$As$_2$ cannot open up a full energy gap between the inverted bulk conduction and valence bands due to the protection of an additional crystallographic symmetry \cite{Dirac_3D} (in the case of Cd$_3$As$_2$ it is the $C_4$ rotational symmetry along the $k_z$ direction \cite{Dai}), which is in contrast to other band-inverted systems such as HgTe \cite{RMP}. This theory predicts \cite{Dai} that the $C_4$ rotational symmetry protects two bulk (3D) Dirac band touching points at two special $\mathbf{k}$ points along the $\Gamma-Z$ momentum space cut-direction, as shown by the red crossings in Fig. 1d. Therefore, Cd$_3$As$_2$ serves a candidate for a spacegroup or crystal structure symmetry protected $C_4$ bulk Dirac semimetal (BDS) phase.

\bigskip
\textbf{Observation of bulk Dirac cone}

%\textbf{Observation of bulk Dirac cone: 3D analog of graphene}

 In order to experimentally identify such a BDS phase, we systematically study the electronic structure of Cd$_3$As$_2$ on the cleaved (001) surface. Fig. 1c shows momentum-integrated ARPES spectral intensity over a wide energy window. Sharp ARPES intensity peaks at binding energies of $E_\textrm{B} \simeq 11$ eV and $41$ eV that correspond to the cadmium $4d$ and the arsenic $3d$ core levels are observed, confirming the chemical composition of our samples. We study the overall electronic structure of the valence band. Fig. 1e shows the second derivative image of an ARPES dispersion map in a 3 eV binding energy window, where the dispersion of several valence bands are identified. Moreover, a low-lying small feature that crosses the Fermi level is observed. In order to resolve it, high-resolution ARPES dispersion measurements are performed in the close vicinity of the Fermi level as shown in Fig. 1f. Remarkably, a linearly dispersive upper Dirac cone is observed at the surface BZ center $\bar{\Gamma}$ point, whose Dirac node is found to locate at a binding energy of $E_{\textrm{B}}\simeq0.2$ eV. At the Fermi level, only the upper Dirac band but no other electronic states are observed. On the other hand, the linearly dispersive lower Dirac cone is found to coexist with another parabolic bulk valence band, which can be seen from Fig. 1e. From the observed steep Dirac dispersion (Fig. 1f), we obtain a surprisingly high Fermi velocity of about 9.8 eV$\cdot$$\AA$ ($\simeq1.5\times10^{6}$ ms$^{-1}$). 
 %$\pm$2)
This is more than 10-fold larger than the theoretical prediction of 0.15 eV$\cdot$ $\AA$  at the corresponding location of the chemical potential \cite{Dai}. Compared to the much-studied 2D Dirac systems, the Fermi velocity of the 3D Dirac fermions in Cd$_3$As$_2$ is thus about 3 times higher than that of in the topological surface states (TSS) of Bi$_2$Se$_3$ \cite{Xia}, $1.5$ times higher than in graphene \cite{Eli} and 30 times higher than that in the topological Kondo insulator phase in SmB$_6$ \cite{SmB6, SmB6_Hasan}. The observed large Fermi velocity of the 3D Dirac band provides clues to understand Cd$_3$As$_2$'s unusually high mobility reported in previous transport experiments \cite{Mobi, Mobi2}. Therefore one can expect to observe unusual magneto-electrical and quantum Hall transport properties under high magnetic field. It is well-known that in graphene the capability to prepare high quality and high mobility samples has enabled the experimental observations of many interesting phenomena that arises from its 2D Dirac fermions. The large Fermi velocity and high mobility in Cd$_3$As$_2$ are among the important experimental criteria to explore the 3D relativistic physics in various Hall phenomena in tailored Cd$_3$As$_2$.

We compare ARPES observations with our theoretical calculations which is qualitatively consistent with previous calculations \cite{Dai}. The reason for the use of our calculations is two fold: first, our calculations are fine tuned based on the characterization of samples used in the present ARPES study, second, sufficiently detailed cuts are not readily available from ref \cite{Dai} which is necessary for a detailed comparison of ARPES data with theory. In theory, there are two 3D Dirac nodes that are expected at two special $\mathbf{k}$ points along the $\Gamma-Z$ momentum space cut-direction, as shown by the red crossings in Fig. 1d. At the (001) surface, these two $\mathbf{k}$ points along the $\Gamma-Z$ axis project on to the $\bar{\Gamma}$ point of the (001) surface BZ (Fig. 1d). Therefore, at the (001) surface, theory predicts one 3D Dirac cone at the BZ center $\bar{\Gamma}$ point, as shown in Fig. 2a. These results are in qualitative agreement with our data, which supports our experimental observation of the 3D BDS phase in Cd$_3$As$_2$. We also study the ARPES measured constant energy contour maps (Fig. 2c and d). At the Fermi level, the constant energy contour consists of a single pocket centered at the $\bar{\Gamma}$ point. With increasing binding energy, the size of the pocket decreases and eventually shrinks to a point (the 3D Dirac point) near $E_{\textrm{B}}\simeq0.2$ eV. The observed anisotropies in the iso-energetic contours are likely due to matrix element effects associated with the standard p-polarization geometry used in our measurements.
\bigskip

\textbf{Three-dimensional dispersive nature}

 A 3D Dirac semimetal is expected to feature nearly linear dispersion along all three momentum space directions close to the crossing point, even though the Fermi/Dirac velocity can vary significantly along different directions. It is well known that in real materials such as pure Bi or graphene or topological insulators the Dirac cones are never perfectly linear over a large energy window yet they can be approximated to be so within a narrow energy window and in comparison to the large effective mass of conventional band electrons in many other materials. In order to probe the 3D nature of the observed low-energy Dirac-like bands in Cd$_3$As$_2$, we performed ARPES measurements as a function of incident photon energy to study the out-of-plane dispersion perpendicular to the (001) surface. Upon varying the photon energy, one can effectively probe the electronic structure at different out-of-plane momentum $k_z$ values in a three-dimensional Brillouin zone and compare with band calculations. In Cd$_3$As$_2$, the electronic structure or band dispersions in the vicinity of its 3D Dirac-like node can be approximated as : $v_{\|}^2(k_x^2+k_y^2)+v_{\perp}^2(k_z-k_0)^2=E^2$, where $k_0$ is the out-of-plane momentum value of the 3D Dirac point. Thus at a fixed $k_z$ value (which is determined by the incident photon energy value), the in-plane electronic dispersion takes the form: $v_{\|}^2(k_x^2+k_y^2)=E^2-v_{\perp}^2(k_z-k_0)^2$. It can be seen that only at $k_z=k_0$ the in-plane dispersion is a gapless Dirac cone, whereas in the case for $k_z\neq{k_0}$ the nonzero $k_z-{k_0}$ term acts as an effective mass term and opens up a gap in the in-plane dispersion relation. Fig. 3a shows the ARPES measured in-plane electronic dispersion at various photon energies. At a photon energy of $102$ eV, a gapless Dirac-like cone is observed, which shows that photon energy $h\nu=102$ eV corresponds to a $k_z$ value that is close to the out-of-plane momentum value of the 3D Dirac node $k_0$. As photon energy is changed away from $102$ eV in either direction, the bulk conduction and valence bands are observed within experimental resolution to be separated along the energy axis and a gap opens in the in-plane dispersion. At photon energies sufficiently away from $102$ eV, such as 90 eV or 114 eV in Fig. 3a, the in-plane gap is large enough so that the bottom of the upper Dirac cone (bulk conduction band) is moved above the Fermi level, and therefore only the lower Dirac cone is observed. We now fix the in-plane momenta at 0 and plot the ARPES data at $k_x=k_y=0$ as a function of incidence photon energy. As shown in Fig. 3b, a $E-k_z$ dispersion is observed in the out-of-plane momentum space cut direction, which is in qualitative agreement with the theoretical calculations (Fig. 3c). The Fermi velocity in the z-direction can be estimated (only at the order of magnitude level) to be about 10$^{5}$ ms$^{-1}$.
 %(3$\pm$2) eV$\cdot$$\AA$. % which is about an order of magnitude smaller than the in-plane Fermi velocity. 
 We note that the sample we used for $k_z$ dispersion measurements (Figs. 3a-c) is relatively $p-$type (Fermi velocity is about 80 meV from the Dirac point) as compared to the sample we used to measure the in-plane dispersion and Fermi surfaces (Figs. 1-2) where chemical potential is about 200 meV from the Dirac point. 
It is important to note that the magnitude of Fermi velocity anisotropy strongly depends on the position of the sample chemical potential ($n-$type sample leads to weaker anisotropy), and therefore the direct comparison between our results and previous transport data in terms of this anisotropy is not applicable.  These systematic incident photon energy dependent measurements show that the observed Dirac-like band disperses along both the in-plane and the out-of-plane directions suggesting its three-dimensional or bulk nature consistent with theory.

In order to further understand the nature of the observed Dirac band, we study the spin polarization or spin texture properties of Cd$_3$As$_2$. As shown in Fig. 3f, spin-resolved ARPES measurements are performed on a relatively $p-$type sample. Two spin-resolved energy-dispersive curve (EDC) cuts are shown at momenta of $\pm0.1$ $\AA^{-1}$ on the opposite sides of the Fermi surface. The obtained spin data shown in Figs. 3g and h show no observable net spin polarization or texture behavior within our experimental resolution, which is in remarkable contrast with the clear spin texture in 2D Dirac fermions on the surfaces of topological insulators. The absence of spin texture in our observed Dirac fermion in Cd$_3$As$_2$ bands is consistent with their bulk origin, which agrees with the theoretical prediction. It also provides a strong evidence that our ARPES signal is mainly due to the bulk Dirac bands on the surface of Cd$_3$As$_2$, whereas the predicted surface (resonance) states \cite{Dai} that lie along the boundary of the bulk Dirac cone projection has a small spectral weight (intensity) contribution to the photoemission signal. In other words, according to our experimental data, the surface electronic structure of Cd$_3$As$_2$ is dominated by the spin-degenerate bulk bands, which is very different from that of the 3D topological insulators.

\bigskip
\bigskip
\textbf{Discussion}
\newline
The distinct semimetal nature of Cd$_3$As$_2$ is better understood from ARPES data if we compare our results with that of the prototype TI, Bi$_2$Se$_3$. In Bi$_2$Se$_3$ as shown in Fig. 4b, the bulk conduction and valence bands are fully separated (gapped), and a linearly dispersive topological surface state is observed that connect across the bulk band-gap. In the case of Cd$_3$As$_2$ (Fig. 4a), there does not exist a full bulk energy gap. On the other hand, the bulk conduction and valence bands ``touch'' (and only ``touch'') at one specific location in the momentum space, which is the 3D band-touching node, thus realizing a 3D BDS. For comparison, we further show that a similar BDS state is also realized by tuning the chemical composition $\delta$ (effectively the spin-orbit coupling strength) to the critical point of a topological phase transition between a normal insulator and a topological insulator. Figs. 4c and d present the surface electronic structure of two other BDS phases in the BiTl(S$_{1-\delta}$Se$_{\delta}$)$_2$ and (Bi$_{1-\delta}$In$_{\delta}$)$_2$Se$_3$ systems. In both systems, it has been shown that tuning the chemical composition $\delta$ can drive the system from a normal insulator state to a topological insulator state \cite{Suyang, Ando, Oh}. The critical compositions for the two topological phase transitions are approximately near $\delta=0.5$ and $\delta=0.04$, respectively. Figs. 4c and d show the ARPES measured surface electronic structure of the critical compositions for both BiTl(S$_{1-\delta}$Se$_{\delta}$)$_2$ and (Bi$_{1-\delta}$In$_{\delta}$)$_2$Se$_3$ systems, which are expected to exhibit the BDS phase. Indeed, the bulk critical compositions where bulk and surface Dirac bands collapse also show Dirac cones with intensities filled inside the cones, which is qualitatively similar to the case in Cd$_3$As$_2$. Currently, the origin of the filling behavior is not fully understood irrespective of the bulk (out-of-plane dispersive behavior) nature of the overall band dispersion interpreted in connection to band calculations (see Fig. 2). Based on the ARPES data in Figs. 4c and d, the Fermi velocity is estimated to be $\sim4$ eV$\cdot\textrm{\AA}$ and $\sim2$ eV$\cdot\textrm{\AA}$ for the 3D Dirac fermions in BiTl(S$_{1-\delta}$Se$_{\delta}$)$_2$ and (Bi$_{1-\delta}$In$_{\delta}$)$_2$Se$_3$ respectively, which is much lower than that of what we observe in Cd$_3$As$_2$, thus likely limiting the carrier mobility. The mobility is also limited by the disorder due to strong chemical alloying. More importantly, the fine control of doping/alloying $\delta$ value and keeping the composition exactly at the bulk critical composition is difficult to achieve \cite{Suyang}, especially while considering the chemical inhomogeneity introduced by the dopants. For example, although similarly high electron mobility on the order of $10^5$ cm$^2$V$^{-1}$s$^{-1}$ has been reported in the bulk states of Pb$_{1-x}$Sn$_x$Se ($x=0.23$) \cite{Ong}, the bulk Dirac fermions there are in fact massive due to the difficulty of controlling the composition exactly at the critical point. These facts taken together exclude the possibility of realizing proposed topological physics including the Weyl semimetal and quantum spin Hall phases using the bulk Dirac states in the Pb$_{1-x}$Sn$_x$Se. These issues do not arise in the stoichiometric Cd$_3$As$_2$ system since its BDS phase is protected by the crystal symmetry, which does not require chemical doping and therefore the natural high electron mobility is retained (not diminished). 
%We note that our crystal of Cd$_3$As$_2$ are observed to be ``stoichiometric within electron probe micro-analyzer (EPMA) and  X-ray diffraction (XRD) resolution limit, but the unavoidable existence of low level defects and its role contributing to the observed phenomenon remains to be explored theoretically". Our sample is not perfectly defect free, however, the global band structure (3D band structure) is not affected by some impurity or defect, which only shift the Fermi level. 
We note that our crystals of  Cd$_3$As$_2$ are nearly stoichiometric within the resolution of electron probe micro-analyzer (EPMA) and  X-ray diffraction (XRD) analysis. The existence of some low level defects is not ruled out. However, these defects do not affect the main conclusion regarding the 3D Dirac band structure ground state of this compound.
Beside Cd$_3$As$_2$ and the topological phase transition critical composition samples as discussed above, we also note that bulk Dirac semimetals unrelated to the combination of $C_4$ symmetry and band-inverted spin-orbit coupling (combination of which has been termed ``topological'' in theory \cite{Dai}) have been studied previously in pnictide BaFe$_2$As$_2$ \cite{Ding}, heavy fermion LaRhIn$_5$ \cite{LaRhIn5}, and organic compound $\alpha$-(BEDT-TTF)$_2$I$_3$ \cite{Organic}. The recent interest is actually focused on spin-orbit based 3D bulk Dirac semimetal phase since the spin-orbit coupling can drive exotic topological phenomena and quantum transport in such materials as the Weyl phases, high temperature linear quantum magnetoresistance and topological magnetic phases \cite{Fang, Ashvin, Balent, 3D_Dirac,Volovik, Dirac_3D, Dirac_semi, Dai}. Our observation of the bulk Dirac states in Cd$_3$As$_2$ provides a unique combination of physical properties, including high spin-orbit coupling strength, high electron mobility, massless nature guaranteed by the crystal symmetry protection without compositional tuning, making it an ideal and unique platform to realize many of the proposed exciting new topological physics \cite{Fang, Ashvin, Balent, 3D_Dirac,Volovik, Dirac_3D, Dirac_semi, Dai}.

In conclusion, we have experimentally discovered the crystalline-symmetry-protected 3D spin-orbit BDS phase in a stoichiometric system Cd$_3$As$_2$ (see Fig. 5). The combination of a large Fermi velocity and very high electron mobility of the 3D carriers with nearly linear dispersion at the crossing point makes it a promising platform to explore novel 3D relativistic physics in various types of quantum Hall phenomena. Our band structure study of the predicted 3D BDS phase also paves the way for designing and realizing a number of related exotic topological phenomena in future experiments. For example, if the $C_4$ crystalline symmetry is broken, the 3D Dirac cone in Cd$_3$As$_2$ can open up a gap and therefore a topological insulator phase is realized in a high mobility setting (current Bi-based TIs feature low carrier mobility). Furthermore upon doping magnetic elements or fabricating superlattice hetero-structures, the 3D Dirac node in Cd$_3$As$_2$ can be split into two topologically protected Weyl nodes, realizing the much sought out Fermi arcs phases in solid-state setting.

%Besides Cd3As2, Na3Bi (refs \cite{Chen} and \cite{SYX}) also feature 3D Dirac phase. However, unlike Cd3As2, Na3Bi is a metastable low mobility compound.

Concurrently posted preprints (refs \cite{MN} (ours) and \cite{Cava}) report ARPES studies of experimental realization of 3D topological Dirac semimetal phase in Cd$_3$As$_2$, however, many of the experimental details and interpretations of the data differ from ours. Later, two other preprints (refs \cite{Chen} and \cite{SYX}) report experimental realization of the 3D Dirac phase in a metastable low mobility compound, Na$_3$Bi.

\bigskip
\bigskip
\textbf{Methods}
\newline

\textbf{Sample growth and characterization} 

Single crystalline samples of Cd$_3$As$_2$ were grown using the standard method, which is described elsewhere \cite{crys_str}.  The Cd$_3$As$_2$ samples used for our ARPES studies show carrier density of  $5.2\times10^{18}$ cm$^{-3}$ and mobility up to $42850$ cm$^2$V$^{-1}$s$^{-1}$ at temperature of 130 K, which is consistent with the mobility of $10^4$ cm$^2$V$^{-1}$s$^{-1}-10^5$ cm$^2$V$^{-1}$s$^{-1}$ reported elsewhere \cite{Mobi, Mobi2}. A slight variation of the value of carrier density and mobility is observed for different growth batch samples. 
We note that our samples show different chemical potential position (measured by ARPES) and different carrier density (measured by transport) depending on the detailed growth conditions.
Moreover, our crystals of  Cd$_3$As$_2$ are nearly stoichiometric within the resolution of electron probe micro-analyzer (EPMA) and  X-ray diffraction (XRD) analysis. 
The existence of some low level defects is not ruled out. 
%However, these defects do not affect the main conclusion regarding the 3D Dirac band structure ground state of this compound.
%We note that our crystal of Cd$_3$As$_2$ are observed to be "stoichiometric within EPMA and XRD resolution limit, but the unavoidable existence of low level defects and its role contributing to the observed phenomenon remains to be explored theoretically".

\textbf{Spectroscopic measurements} 

ARPES measurements for the low energy electronic structure were performed at the PGM beamline in Synchrotron Radiation Center (SRC) in Wisconsin, and at the beamlines 4.0.3, 10.0.1 and 12.0.1 at the Advanced Light Source (ALS) in Berkeley California, equipped with high efficiency VG-Scienta R4000 or R8000 electron analyzers. Spin-resolved ARPES measurements were performed at the ESPRESSO endstation at HiSOR. Photoelectrons are excited by an unpolarized He-I$\alpha$ light (21.21 eV). The spin polarization is detected by state-of-the-art very low energy electron diffraction (VLEED) spin detectors utilizing preoxidized Fe(001)-p($1 \times 1$)-O targets \cite{Okuda_BL9B}. The two spin detectors are placed at an angle of 90$^\circ$ and are directly attached to a VG-Scienta R4000 hemispheric analyzer, enabling simultaneous spin-resolved ARPES measurements for all three spin components as well as high resolution spin integrated ARPES experiments. The energy and momentum resolution was better than 40 meV and $1\%$ of the surface BZ for spin-integrated ARPES measurements at the SRC and the ALS, and 80 meV and $3\%$ of the surface BZ for spin-resolved ARPES measurements at ESPRESSO endstation at HiSOR. Samples were cleaved \textit{in situ} and measured at $10-80$ K in a vacuum better than $1\times10^{-10}$ torr. They were found to be very stable and without degradation for the typical measurement period of 20 hours.

\textbf{Theoretical calculations} 

The first-principles calculations are based on the generalized gradient approximation (GGA) \cite{Perdew} using the projector augmented wave method \cite{Blochl,Blochl_1} as implemented in the VASP package \cite{Kress, Kress_1}. The experimental crystal structure was used \cite{crys_str}. The electronic structure calculations were performed over $4\times4\times2$ Monkhorst-Pack k-mesh with the spin-orbit coupling included self-consistently.

\bigskip
\bigskip
\textbf{Acknowledgements}
\newline
The work at Princeton and Princeton-led synchrotron X-ray-based measurements and the related theory at Northeastern University are supported by the Office of Basic Energy Sciences, US Department of Energy (grants DE-FG-02-05ER46200, AC03-76SF00098 and DE-FG02-07ER46352). 
We thank J. Denlinger, S.-K. Mo and A. Fedorov for beamline assistance at the DOE supported  Advanced Light Source (ALS-LBNL) in Berkeley. We also thank M. Bissen and M. Severson for beamline assistance at SRC, WI.
M.Z.H. acknowledges Visiting Scientist support from LBNL, Princeton University and the A. P. Sloan Foundation.

\bigskip

\textbf{Author contributions}
\newline
M.N., and S.-Y.X. performed the experiments with assistance from N.A., G.B., C.L., I. B. and M.Z.H.; 
M.N., and M.Z.H. performed data analysis, figure planning and draft preparation;
R. S. and F.-C. C. provided the single-crystal samples and performed sample characterization;
T.R.C., H.T.J., H.L., and A.B. carried out calculations; M.Z.H. was responsible for the conception and the overall direction, planning and integration among different research units.

\textbf{Additional information}
%\newline
%Supplementary Information accompanies this paper is available at http://www.nature.com naturecommunications.

Competing financial interests: The authors declare no competing financial interests.
\newline
\*Correspondence and requests for materials should be addressed to
\newline
M.Z.H. (Email: mzhasan@princeton.edu).

\newpage

\noindent

\newpage

\begin{figure*}
\centering
\includegraphics[width=15cm]{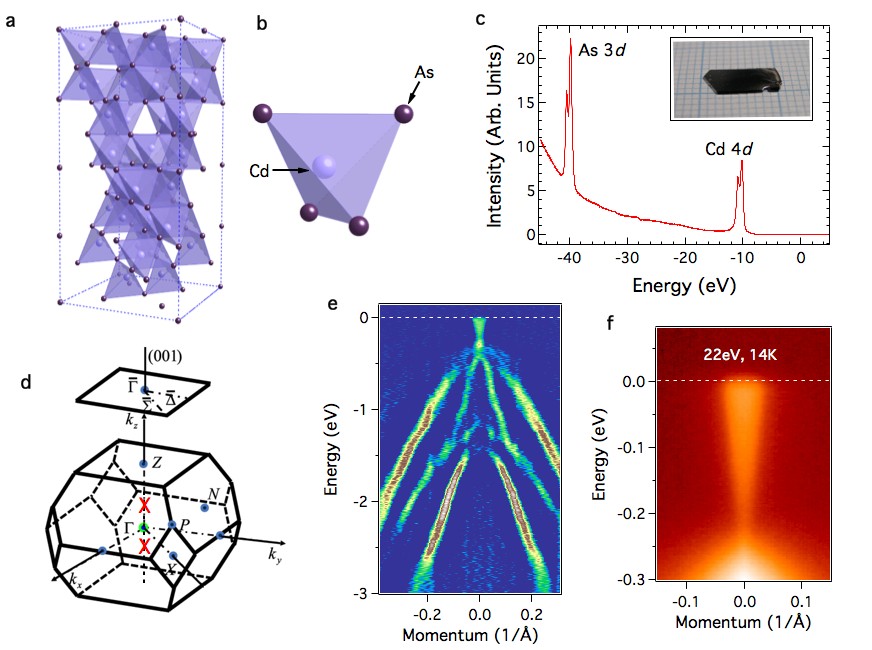}
\caption{\textbf{Brillouin zone symmetry and 3D Dirac cone.} \textbf{a,} Cd$_3$As$_2$ crystalizes in a tetragonal body center structure with space group of $I4_1$cd, which has 32 number of formula units in the unit cell. The tetragonal structure has lattice constant of  $a= 12.670$ \AA, $b=12.670$ \AA, and $c=25.480$ \AA. \textbf{b,} The basic structure unit is a 4 corner-sharing CdAs$_3$-trigonal pyramid. \textbf{c,} Core-level spectroscopic measurement where Cd 4$d$ and As 3$d$ peaks are clearly observed. Inset shows a picture of the Cd$_3$As$_2$ samples used for ARPES measurements. The flat and mirror-like surface indicates the high quality of our samples. \textbf{d,} The bulk Brillouin zone (BZ) and the projected surface BZ along the (001) direction. The red crossings locate at $(k_x,k_y,k_z)=(0,0,0.15\frac{2\pi}{c*})$ ($c*=c/a$). They denote the two special $\mathbf{k}$ points along the $\Gamma-Z$ momentum space cut-direction, where 3D Dirac band-touchings are protected by the crystalline $C_4$ symmetry along the $k_z$ axis.   \textbf{e,} Second derivative image of ARPES dispersion map of Cd$_3$As$_2$ over the wider binding energy range. Various bands are well-resolved up to 3 eV binding energy range. \textbf{f,} ARPES $E_{\textrm{B}}-k_x$ cut of Cd$_3$As$_2$ near the Fermi level at around surface BZ center $\bar{\Gamma}$ point.}
\end{figure*}

\begin{figure*}
\centering
\includegraphics[width=17cm]{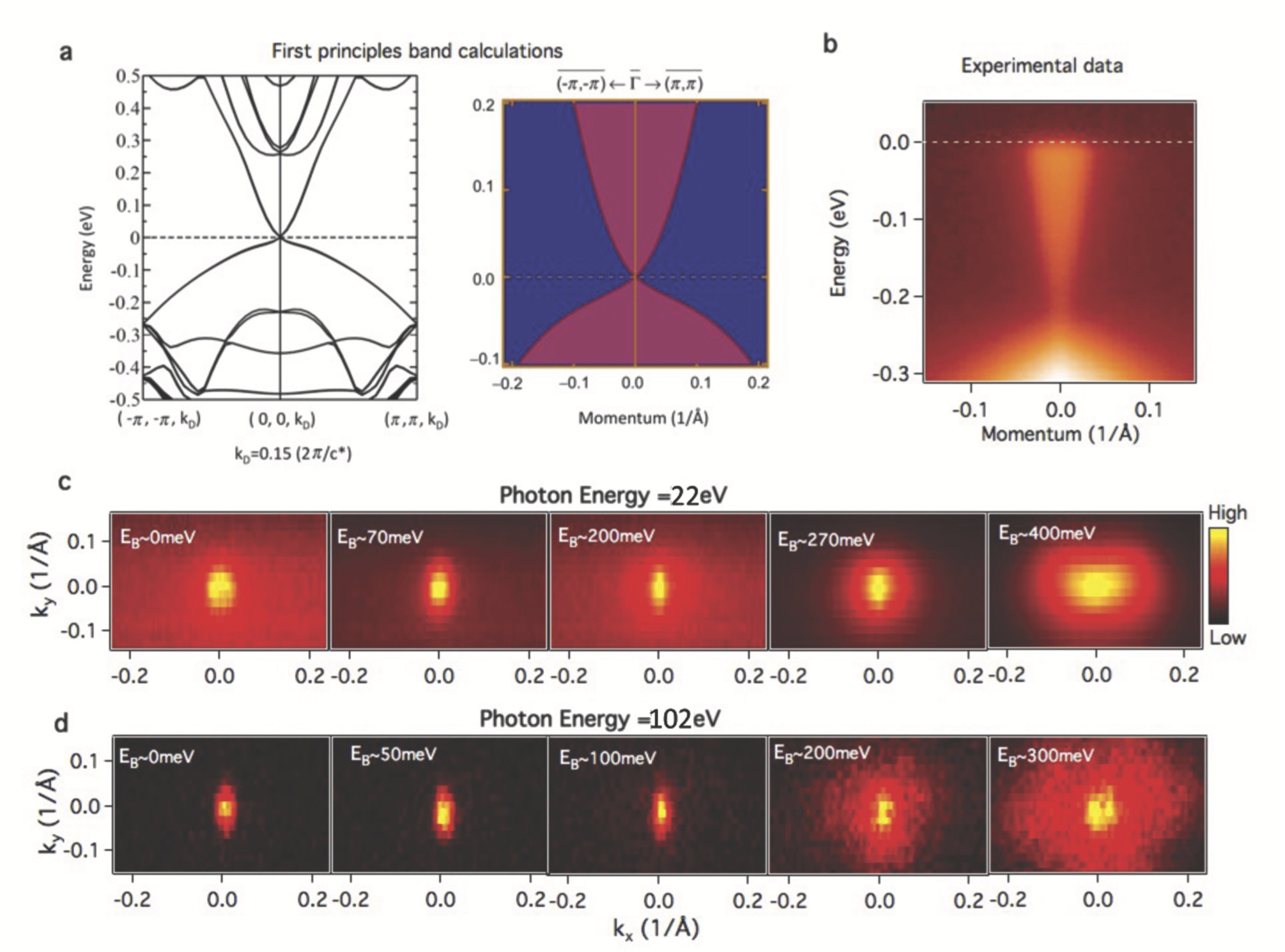}
\caption{\textbf{Observation of in-plane dispersion in Cd$_3$As$_2$.} \textbf{a,} Left: First principles calculation of the bulk electronic structure along the $(\pi,\pi,0.15\frac{2\pi}{c*})-(0,0,0.15\frac{2\pi}{c*})$ direction ($c*=c/a$). Right: Projected bulk band structure on to the (001) surface, where the shaded area shows the projection of the bulk bands. \textbf{b,} ARPES measured dispersion map of Cd$_3$As$_2$, measured with photon energy of 22 eV and temperature of 15 K along the $(-\pi,-\pi)-(0,0)-(\pi,\pi)$ momentum space cut direction. \textbf{c,} ARPES constant energy contour maps using photon energy of 22 eV on Cd$_3$As$_2$ growth batch I. \textbf{d,}  ARPES constant energy contour maps using photon energy of 102 eV on Cd$_3$As$_2$ batch II. In order to achieve chemical potential (carrier concentration) control, we have prepared different batches of samples under slightly different growth conditions (temperature and growth time). For the two batches studied here, batch I is found to be slightly more $n-$type than batch II (e.g. compare batch I in Fig. 1f with batch II in Fig. 3a rightmost panel). }
\end{figure*}

\newpage

\clearpage
\begin{figure}
\centering
\includegraphics[width=15cm]{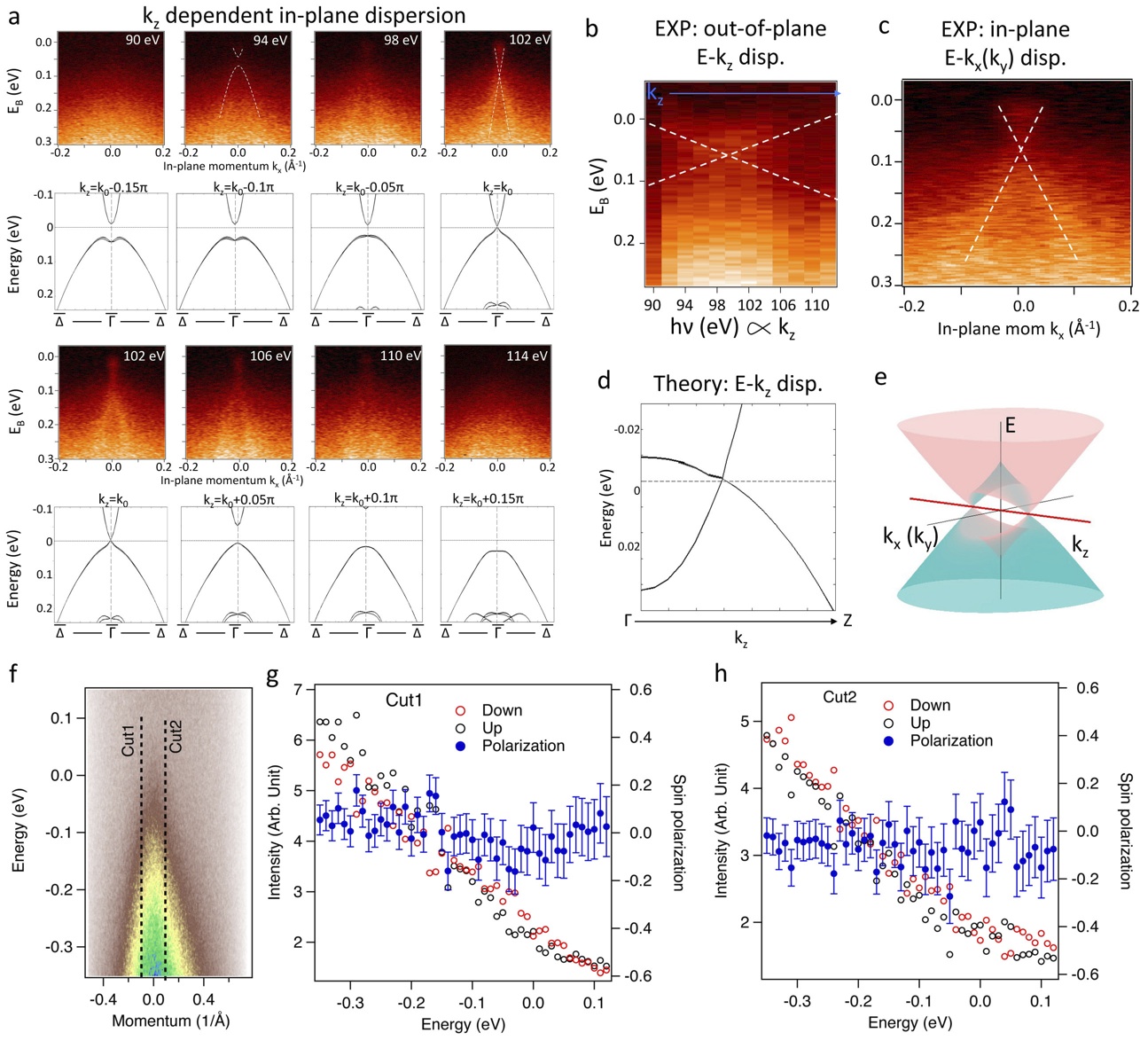}
\caption{\textbf{Observation of out-of-plane dispersion in Cd$_3$As$_2$.} \textbf{a,} ARPES dispersion maps at various incident photon energies are shown in the first and third rows. First principle calculated in-plane electronic dispersion at different $k_z$ values near the 3D Dirac node $k_0$ is plotted in the second and forth rows. \textbf{b,} ARPES measured out-of-plane linear $E-k_z$ dispersion. \textbf{b,} ARPES measured in-plane $E-k_x$ dispersion. The white dotted lines are guides to the eye tracking the out-of-plane dispersion.  \textbf{d,} Theoretically calculated out-of-plane $E-k_z$ dispersion near the 3D Dirac node shown over a wider energy window. \textbf{e,} Schematic (cartoon) of the 3D (anisotropic) Dirac semimetal band structure in Cd$_3$As$_2$. \textbf{f,} Spin-integrated ARPES dispersion cut measured on the sample used for spin-resolved measurements. The dotted lines indicate the momentum locations for the spin-resolved EDC cuts. \textbf{g and h,} Spin-resolved ARPES intensity (black and red circles) and measured net spin polarization (blue dots) for Cuts 1 and 2. Error bars represent the experimental uncertainties in determining the spin polarization.}
\end{figure}

\begin{figure}
\centering
\includegraphics[width=16cm]{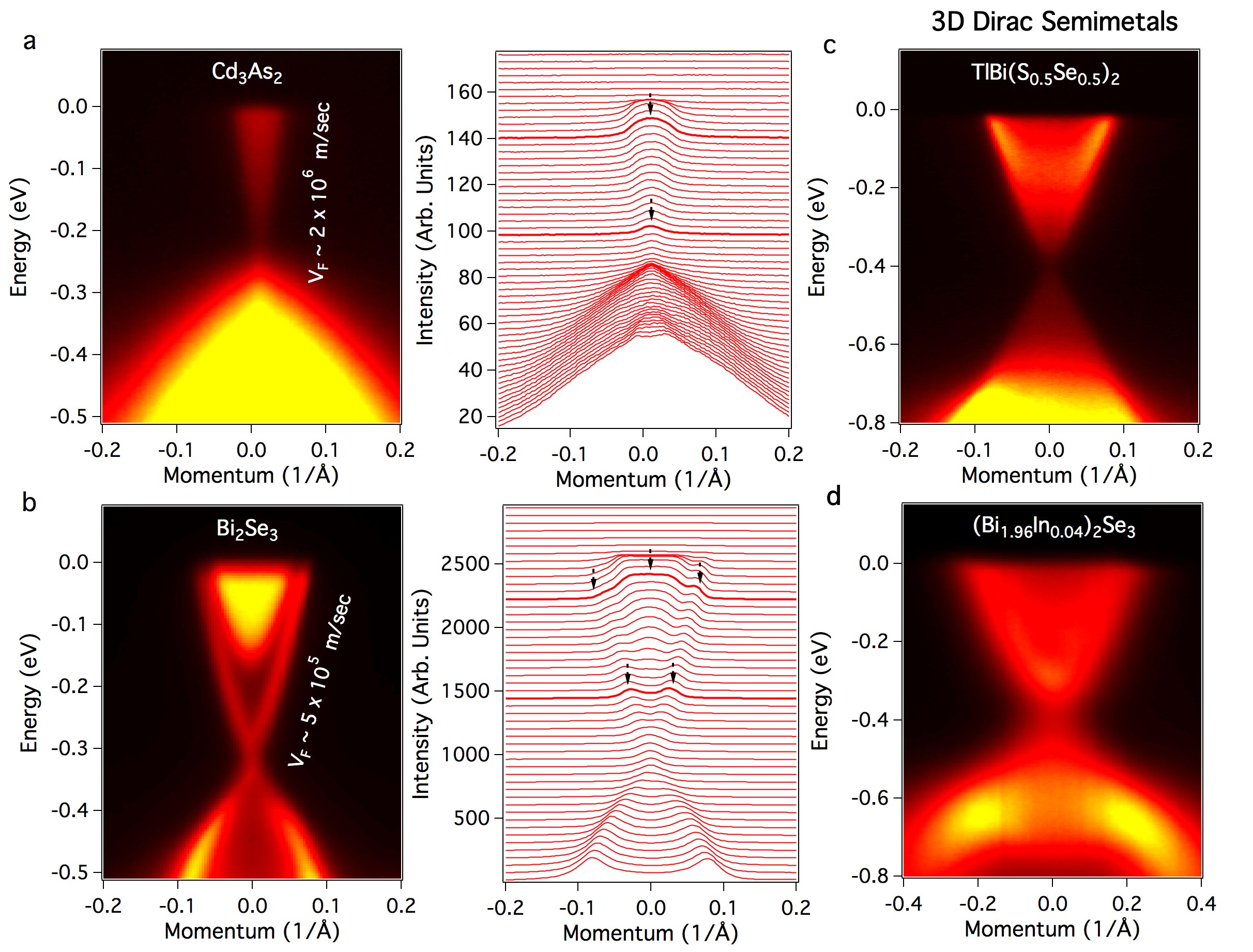}
\caption{\textbf{Surface electronic structure of 2D and 3D Dirac fermions.} \textbf{a,} ARPES measured surface electronic structure dispersion map of Cd$_3$As$_2$ and its corresponding momentum distribution curves (MDCs). \textbf{b,} ARPES measured surface dispersion map of the prototype TI Bi$_2$Se$_3$ and its corresponding momentum distribution curves. Both spectra are measured with photon energy of 22 eV and at a sample temperature of 15 K. The black arrows show the ARPES intensity peaks in the MDC plots. \textbf{c and d} ARPES spectra of two Bi-based 3D Dirac semimetals, which are realized by fine tuning the chemical composition to the critical point of a topological phase transition between a normal insulator and a TI: \textbf{c,} TlBi(S$_{1-\delta}$Se$_\delta$)$_2$ ($\delta=0.5$) (see Xu $et$ $al.$ \cite{Suyang}), and (Bi$_{1-\delta}$In$_{\delta}$)$_2$Se$_3$ ($\delta=0.04$) (see Brahlek $et$ $al.$ \cite{Oh}). \textbf{d,}. Spectrum in panel \textbf{c} is measured with photon energy of 16 eV and spectrum in panel \textbf{d} is measured with photon energy of 41 eV. For the 2D topological surface Dirac cone in Bi$_2$Se$_3$, a distinct in-plane ($E_{\textrm{B}}-k_x$) dispersion is observed in ARPES, whereas for the 3D bulk Dirac cones in Cd$_3$As$_2$, TlBi(S$_{0.5}$Se$_{0.5}$)$_2$, and (Bi$_{0.96}$In$_{0.04}$)$_{2}$Se$_3$, a Dirac-cone-like intensity continuum is also observed. }
\end{figure}

\begin{figure}
\centering
\includegraphics[width=15cm]{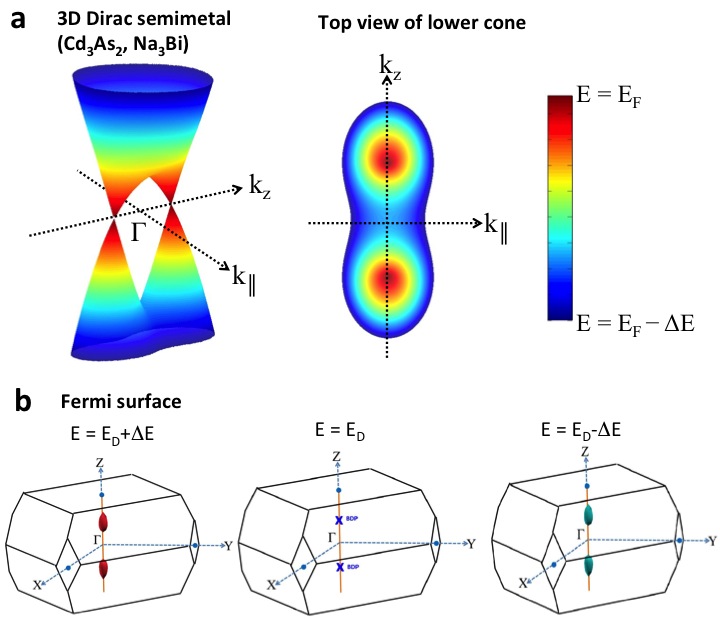}
\caption{\textbf{Essence of 3D Dirac semimetal phase}. \textbf{a,} Cartoon view of dispersion of 3D Dirac semimetal. \textbf{b,} Schematic view of the Fermi surface above the Dirac point (left panel), at the Dirac point (middle panel) and below the Dirac point (right panel). }
\end{figure}

%(Cd$_3$As$_2$ (see \cite{MN, Cava}) and Na$_3$Bi (see \cite{Chen, SYX}))

%
%\begin{figure*}
%\centering
%\includegraphics[width=17cm]{Fig5}
%\caption{\textbf{Photon energy dependent measurements of other 3D Dirac semimetals.} Photon energy dependent ARPES spectra are presented for two 3D Dirac semimetals: \textbf{a,} TlBi(S$_{0.5}$Se$_{0.5}$)$_2$, and  \textbf{b,} (Bi$_{0.96}$In$_{0.04}$)$_{2}$Se$_3$. The photon energies used to measure the spectrum are noted on each plot.}
%\end{figure*}

\end{document}